\newcommand{\be}{\begin{equation}}\newcommand{\ee}{\end{equation}}
\newcommand{\bea}{\begin{eqnarray}}\newcommand{\eea}{\end{eqnarray}}
\newcommand{\pa}{\partial}
\newcommand{\p}[1]{(\ref{#1})}
\newcommand{\lpr}{{\cal L}^{+4}}

\newcommand{\fracds}[2]{\displaystyle \frac{#1}{#2} }
\documentstyle[12pt]{article}

\def\a{\alpha} \def\b{\beta}\def\g{\gamma} \def\e{\epsilon}
\def\d{\delta}
\def\G{\Gamma} \def\O{\Omega}
\begin{document}
\thispagestyle{empty}
\begin{flushright}   JHU-TIPAC-920022\\
MPI-Ph/92-84\\ October 1992
\end{flushright}
\bigskip\bigskip\begin{center} {\bf
\Huge{Harmonic potentials for quaternionic symmetric $\sigma$-models}}
\end{center}  \vskip 1.0truecm
\centerline{\bf A. Galperin${}^{(a)\dag}$
and O. Ogievetsky${}^{(b)\ddag}$}
\vskip5mm
\centerline{${}^{(a)}$ Department of Physics and Astronomy}
\centerline{Johns Hopkins University, Baltimore, MD 21218, USA}
\vskip5mm
\centerline{${}^{(b)}$Max-Planck-Institut f\"{u}r
 Physik, Werner-Heisenberg-Institut}
\centerline{F\"ohringer Ring 6, 8000 Munich 40, West Germany}

\vskip 4cm
\bigskip \nopagebreak \begin{abstract}
We construct $N=2$ superspace Lagrangians for quaternionic symmetric
$\sigma$-models $ G/H\times Sp(1)$, or equivalently,
quaternionic potentials for these symmetric spaces.
They are
homogeneous $H$ invariant polynomials of order 4 which are
similar to the
quadratic Casimir operator of $H$. The construction is based
on an identity for the structure constants specific for quaternionic
symmetric spaces.
\end{abstract}
\vskip 2.4cm
\bigskip \nopagebreak \begin{flushleft} \rule{2
in}{0.03cm} \\ {\footnotesize \ ${}^{\dag}$  On leave from the Laboratory
of Theoretical Physics, JINR, Dubna, Russia}
\\ {\footnotesize \ ${}^{\ddag}$ On leave from P.N. Lebedev Physical
 Institute, Theoretical Department, 117924 Moscow, Leninsky
 prospect 53, Russia}
 \end{flushleft}

\newpage\setcounter{page}1

\section{Introduction}

Quaternionic spaces naturally appear in physics as solutions to
self-dual gravity with a cosmological constant \cite{egh}
as well as target spaces of
$\sigma$-models in  $N=2$ matter couplings to $N=2$
supergravity \cite{baggwitt}. A general description of these
$\sigma$-models in $N=2$ superspace was developed in \cite{bagger}.
It is based on the notion of harmonic superspace \cite{gikos}; for a
detailed exposition of this approach to
quaternionic manifolds see \cite{gio}.
In this description all the information about the quaternionic metric on
the target space is contained in the quaternionic potential ${\cal L}^{+4}$.
The latter is an analog of the K\"ahler potential for K\"ahler manifolds
relevant to $N=1$ supersymmetry \cite{zu}.

The $N=2$ superspace action for the general $4n$-dimensional
quaternionic $\sigma$-model has the following form
\be S=\int d\zeta^{-4}du \{ Q_\alpha^+D^{++}Q^{+\alpha} -q^+_i D^{++}
q^{+i} + \lpr (Q^+,q^+,u^-)\}.\label{i1}\ee
Here integration is performed over the $N=2$ analytic
superspace with coordinates $\zeta, u^{\pm}_i$,
$Q^+_\alpha(\zeta, u), \; \alpha =1,...,2n$
and $q^+_i(\zeta, u), \;i=1,2$ are analytic $N=2$ superfields,
$u^\pm_i,i=1,2$ are the
$SU(2)/U(1)$ isospinor harmonics, $u^{+i}u^-_i=1$.
Finally, $D^{++}$ is a supercovariant
derivative
with respect to harmonics. Its only property we will need
is $D^{++}u^-_i=u^+_i$
(for details and notation see \cite{gikos},\cite{bagger}).

The quaternionic potential $\lpr$
is homogeneous in $Q^+,q^+$ of degree two, it carries $U(1)$-charge
$+4$, does not depend on $u^+$ and otherwise is an arbitrary
real function. The reality is defined by the ordinary complex conjugation
combined with the involution $u^+_i \mapsto u^-_i\ , u^-_i\mapsto -u^+_i$.
The action \p{i1} bears a remarkable analogy to the Hamiltonian mechanics
\cite{go} with the harmonic derivative $D^{++}$
similar to time derivative, $Q^+, q^+$ to phase space coordinates
and the Poisson brackets given by
\be
\{f,g\}^{--} = {1\over 2}\Omega^{\alpha\beta}{\pa f\over \pa Q^{+\alpha}}
{\pa g\over \pa Q^{+\beta}}- {1\over 2}\epsilon^{ij}{\pa f\over\pa q^{+i}}
{\pa g\over\pa q^{+j}},
\label{pb}\ee
where $\Omega^{\alpha\beta}$ and $\epsilon^{ij}$ are the invariant
antisymmetric $Sp(n)$ and $Sp(1)$ tensors, respectively, and $Q^{+\alpha}=
\Omega^{\alpha\beta}Q^+_\beta$, $q^{+i}=\epsilon^{ij}q^+_j$.
The quaternionic potential $\lpr$ is quite analogous to the Hamiltonian
in mechanics; we will call $\lpr$ Hamiltonian, too.

Isometries of the $\sigma$-model \p{i1}
are generated by Killing potentials
$K^{++}_A(Q^+, q^+, u^-)$ which should obey the conservation law
\be
\pa^{++}K^{++}_A +\{K^{++}_A, \lpr\}^{--}=0\ .
\label{cons}\ee
Under the Poisson brackets
\p{pb}, the Killing potentials form the Lie algebra of the
isometry group,
\be
\{K^{++}_A, K^{++}_B\}^{--}=f_{AB}^{\ \ \ C}K^{++}_C\ .\label{Lie}
\ee
Here $\partial_{++}$ is a partial derivative with respect to harmonics,
$\partial_{++}u^-_i=u^+_i$.

An interesting problem is to identify Hamiltonians $\lpr$
for known quaternionic manifolds.
The simplest  among the latter are quaternionic symmetric
spaces described in \cite{wolf}.
There is precisely one compact and one noncompact
quaternionic cosets for each simple complex Lie group.
In $N=2$ supergravity one encounters the noncompact versions of these:
\be\begin{array}{cccc}
\fracds{SU(n,2)}{U(n) \times Sp(1)}&\
\fracds{SO(n,4)}{SO(n)\times SU(2) \times Sp(1)}&  \
\fracds{Sp(n,1)}{Sp(n)\times Sp(1)} &
\fracds{G_{2(+2)}}{SU(2)\times Sp(1)}\\[3ex]
\fracds{F_{4(+4)}}{Sp(3)\times Sp(1)}&\
\fracds{E_{6(+2)}}{SU(6)\times Sp(1)}&\
\fracds{E_{7(-5)}}{SO(12)\times Sp(1)}&
\fracds{E_{8(-24)}}{E_7 \times Sp(1)}.
\end{array}
\label{one}
\ee
They have the form  $G/H\times Sp(1)$ with $H \in Sp(n)$.
The number in brackets for the exceptional groups
refers to the choice of a real form of the group. It is equal to the
difference
between the numbers of noncompact and compact generators for
this real form. The compact case corresponds to the compact real form of
the numerator.

In the present Letter we construct the Hamiltonians $\lpr$
for all these spaces (some of them have been found in \cite{bagger}).
We find that they have
an interesting universal algebraic meaning. To describe it, consider a
hamiltonian action of a compact Lie algebra $H$ on a symplectic manifold
$M$. Denote generators of $H$ by $\Gamma_a$ and
functions on $M$ corresponding to $\Gamma_a$ by $K_a$.
Let $g_{ab}$ be an invariant nondegenerate metric on $H$.
Then one can form a function ${\cal L} =g^{ab}K_aK_b$, which we will
call a Casimir function for its similarity with the Casimir operator.
In the case of the quaternionic symmetric space $G/H\times Sp(1)$
(which is coordinatized by the set $Q^+, q^+, u^-$) every
generator $\G_A$ of $G$ maps to a function $K^{++}_A(Q^+, q^+, u^-)$,
and the action of $K^{++}$ is
given by the Poisson brackets \p{pb} with $K^{++}$.
The restriction of the
Killing--Cartan metric of $G$ to $H$ is nondegenerate. We prove that
the potentials for the quaternionic symmetric spaces are given by
\be
\lpr={(n+2)\over 3(q^{+i}u^-_i)^2} g^{ab}K^{++}_aK^{++}_b \ ,\label{l}
\ee
where  $K^{++}_a$ generate the subgroup
$H$ from the denominator and $4n$ is the space dimension.
 The quaternionic potentials $\lpr$ are the same in compact
and noncompact cases. The only difference is the relative sign of
``kinetic'' terms for $Q^+$ and $q^+$ in the action \p{i1} and in the
Poisson brackets \p{pb}.

A direct way to prove this statement would be to follow the general
procedure \cite{gio} which
establishes a one-to-one correspondence between
the quaternionic spaces and the quaternionic potentials $\lpr$.
However, in the case of symmetric spaces the existence of
of conserved Killing potentials for the isometry group
$G$ turns out to
be sufficient to determine the potential $\lpr$.
We find  $\lpr$
and the set of functions $K^{++}_{ij}, K^{++}_{\alpha i}$ and $K^{++}_a$
such that all $K^{++}$ are conserved in the
sense \p{cons}, $K^{++}_{ij}$ form the Lie algebra of
$Sp(1)$, $K^{++}_a$ form the Lie algebra of
$H$, and together
with $K^{++}_{\alpha i}$ they all form the Lie algebra of
$G$ -- all this with respect to the Poisson brackets \p{pb}.
The solution to these requirements
defines locally a quaternionic manifold
with an action of $G$ on it. By the
construction, we will see that $H$ and $Sp(1)$ are stability
subgroups, and the whole
space is coordinatized
by the coset parameters. Therefore this quaternionic
space can be nothing else
but the quaternionic symmetric space
$ G/H\times Sp(1)$.

\section{An identity for quaternionic symmetric
spaces}

As mentioned in the Introduction every
complex simple Lie group $G_c$
has two real forms  $G$ which contain
such a subgroup $H\times Sp(1)$
that the coset space $G/H\times Sp(1)$ is
quaternionic and symmetric.
In this section we consider the Jacobi identities which result from the
decomposition of $G$ with respect to $H\times Sp(1)$
and derive a general identity,
which plays a central role in our construction.

We start with writing the Lie algebra of $G$:
\be [\Gamma_{\alpha i},\Gamma_{\beta j}] =\epsilon_{ij}
t^{\ \ \ a}_{\alpha\beta}\Gamma_a +\Omega_{\alpha\beta}\Gamma_{ij},
\label{2.5}\ee
\be [\Gamma_a ,\Gamma_{\alpha i}]=t_{a\alpha}^{\ \ \; \beta}
\Gamma_{\beta i},\label{2.6}\ee
\be
[\Gamma_a ,\Gamma_b ]=f_{ab}^{\ \ c}\Gamma_c ,\label{alg}\ee
\be
[\Gamma_{ij},\Gamma_{\alpha k}]=\epsilon_{ik}\Gamma_{\alpha j}
+\epsilon_{jk}\Gamma_{\alpha i}, \;\;\;
[\Gamma_{ij},\Gamma_a]=0\label{2.8}\ ,\ee
\be
[\Gamma_{kl},\Gamma_{ij}]=\epsilon_{ki}\Gamma_{lj}+
\epsilon_{li}\Gamma_{kj}+\epsilon_{kj}\Gamma_{li}
+\epsilon_{lj}\Gamma_{ki},\label{2.9}\ee
where $\Gamma_a$, $\Gamma_{ij}$ and $\Gamma_{\alpha i}$ are the generators
corresponding to the stability subgroups $H$ and $Sp(1)$ and the
coset space $G/H\times Sp(1)$, respectively.

For these commutators to form a Lie
algebra a number of Jacobi identities has to be satisfied. The nontrivial
identities occur for the following triples :
$ (\Gamma_{\alpha i}, \Gamma_{\beta j}, \Gamma_{\gamma k})$,
$(\Gamma_{\alpha i}, \Gamma_{\beta j}, \Gamma_a)$,
$(\Gamma_{\alpha i}, \Gamma_a, \Gamma_b)$.
The first Jacobi identity reads
$$
\epsilon_{ij}t_{\alpha\beta}^{\ \ \ a}
t_{a\gamma}^{\ \ \sigma}\Gamma_{\sigma k} +
\Omega_{\alpha\beta}(\epsilon_{ik}\Gamma_{\gamma
j}+\epsilon_{jk}\Gamma_{\gamma i}) +
$$
\be
\epsilon_{jk}t_{\beta\gamma}^{\ \ \ a}
t_{a\alpha}^{\ \ \sigma}\Gamma_{\sigma i} +
\Omega_{\beta\gamma}(\epsilon_{ji}\Gamma_{\alpha
k}+\epsilon_{ki}\Gamma_{\alpha j})+
\label{1}\ee
$$
\epsilon_{ki}t_{\gamma\alpha}^{\ \ \ a}
t_{a\beta}^{\ \ \sigma}\Gamma_{\sigma j} +
\Omega_{\gamma\alpha}(\epsilon_{kj}\Gamma_{\beta
i}+\epsilon_{ij}\Gamma_{\beta k})=0.
$$
Symmetrizing it in $i,j$ we obtain
\be
\Pi_{\gamma\beta\alpha}^{\ \ \ \ \sigma} -\Pi_{\gamma\alpha\beta}^{\ \ \ \
\sigma} =
\Omega_{\beta\gamma}\d_\alpha^\sigma+
\Omega_{\gamma\alpha}\d_\beta^\sigma -
2\Omega_{\alpha\beta}\d_\gamma^\sigma\ ,
\label{z2}\ee
where
\be \Pi_{\a\b\g}^{\ \ \ \ \sigma}=
t_{\a\b}^{\ \ \ a}t_{a\g}^{\ \ \ \sigma}\ .\label{z4}\ee
Antisymmetrization gives
      \be
2\Pi_{\a\b\g}^{\ \ \ \ \sigma}- \Pi_{\beta\gamma\alpha}^{\ \ \ \ \sigma}
-\Pi_{\gamma\alpha\beta}^{\ \ \ \ \sigma} =
3(\Omega_{\beta\gamma}\d_\alpha^\sigma-
\Omega_{\gamma\alpha}\d_\beta^\sigma),
\label{z3}\ee
which is a consequence of (\ref{z2}), and therefore does not produce
new restrictions.

The Jacobi identities for the second and third triples
relate the structure
constants $f_{ab}^{\ \ c}$ with $t_{\a\b}^{\ \ \ a}$ and
$t_{a\alpha}^{\ \ \ \beta}$:
\be
t_{\alpha\beta}^{\ \ a}f_{ab}^{\ \ c}+
t_{\g\beta}^{\ \ \ c} t_{b\a}^{\ \ \ \g}
+ t_{\g\a}^{\ \ \ c}t_{b\b}^{\ \ \ \g}=0\ ,
\label{z5}\ee
and
\be
t_{c\a}^{\ \ \ \g}f_{ab}^{\ \ c}+t_{a\a}^{\ \ \ \b} t_{b\b}^{\ \ \ \g}
- t_{b\a}^{\ \ \ \b}t_{a\b}^{\ \ \ \g}=0\ .
\label{z5a}\ee
The Jacobi identity for the triple
$(\Gamma_{\alpha i}, \Gamma_{\beta j}, \Gamma_a)$ gives
also
\be t_{b\a}^{\ \ \ \g}\O_{\g\b}=
 t_{b\b}^{\ \ \ \g}\O_{\g\a}\ ,\label{z6}\ee
which says that $\G_a$ generate a subgroup of a symplectic group
preserving $\O_{\a\b}$.

The Jacobi identity for the structure constants $f_{ab}^{\ \ \ c}$
follows from the Jacobi identities above.

Using \p{z2} and \p{z6} one finds
\be \Pi_{\g\b\a}^{\ \ \ \ \b}=(2n+1)\O_{\g\a}\ .\label{z9}\ee

We will need some information about the Killing metric
for $G$.
The involution $\Gamma_\alpha\mapsto -\Gamma_\alpha$,
$\Gamma_a\mapsto \Gamma_a$ and $\Gamma_{ij}\mapsto \Gamma_{ij}$,
entering the definition of symmetric spaces, leaves the Killing
metric $g$ of the group $G$ invariant. Therefore the subspace
$\{ \Gamma_a ,\Gamma_{ij}\}$ is orthogonal to the subspace
$\{ \Gamma_\alpha\}$. The subspaces $\{ \Gamma_a\}$ and $\{
\Gamma_{ij}\}$ are orthogonal as well.
In other words the Killing metric is block-diagonal.
The components
$g_{ab}$ (for the generators of $H$) and
$g_{\a i,\b j}$ (for the coset generators)
of the Killing metric are
\be
g_{ab}=2t_{a\alpha}^{\ \ \ \beta}t_{b\beta}^{\ \ \ \alpha}+
f_{ac}^{\ \ d}f_{bd}^{\ \ \ c},
\label{ab}\ee
\be
g_{\alpha i,\beta j}=-4(n+2)\epsilon_{ij}\Omega_{\alpha\beta}\ .
\label{kos}\ee
Note that the restriction of the Killing metric for $G$ to $H$
is nondegenerate and
might not coincide with the Killing metric for $H$: the subgroup
$H$ may contain more than one simple factor, or  a $U(1)$ factor.

With the help of eq-ns \p{ab}
and \p{kos} the antisymmetry of the structure
constants implies
\be
g_{ab}t_{\alpha\beta}^{\ \ \ b}=
4(n+2)\Omega_{\beta\gamma}t_{a\alpha}^{\ \ \ \gamma} \ .\label{anti}
\ee

Define a quantity
\be \Pi_{\a\b\g\d}=\O_{\d\sigma}\Pi_{\a\b\g}^{\ \ \ \ \sigma}
\ .\label{z11}\ee
It is symmetric in $\a ,\b$ and in $\g ,\d$. Moreover, (\ref{anti})
implies that it is invariant under $\a\b\leftrightarrow
\g\d$ as well. Therefore the totally symmetric part
$S_{\a\b\g\d}$ of
$\Pi_{\a\b\g\d}$ is
$S_{\a\b\g\d}=$ $(\Pi_{\a\b\g\d}+\Pi_{\a\g\b\d}+\Pi_{\a\d\g\b})/3$.
It follows from (\ref{z2}) that $S$ can be written in the form
\be S_{\a\b\g\d}=\Pi_{\a\b\g\d}+\O_{\a\g}\O_{\b\d}+\O_{\a\d}\O_{\b\g}
\ .\label{z12}\ee

Now we are going to derive an identity for the tensor $ S_{\a\b\g\d}$. To
this end we multiply \p{z5a} by $t_{\mu\nu}^{\ \ \ a}t_{\sigma\tau}^{\
\ \ b}$ and use \p{z5} to eliminate the structure constants $f_{ab}^{\ \
c}$:
\be
\Pi_{\rho\mu\alpha}^{\ \ \ \gamma}\Pi_{\sigma\tau\nu}^{\ \ \ \rho}+
\Pi_{\rho\nu\alpha}^{\ \ \ \gamma}\Pi_{\sigma\tau\mu}^{\ \ \ \rho} +
\Pi_{\mu\nu\rho}^{\ \ \ \gamma}\Pi_{\sigma\tau\alpha}^{\ \ \ \rho}-
\Pi_{\sigma\tau\rho}^{\ \ \ \gamma}\Pi_{\mu\nu\alpha}^{\ \ \ \rho} =0.
\label{z13}\ee
This identity has the following algebraic meaning. One can find the
structure constants $f_{ab}^{\ \ \ c}$
using either of eq-ns (\ref{z5}) or (\ref{z5a}). Therefore
there is a compatibility condition ensuring that we obtain the
same $f_{ab}^{\ \ \ c}$, which is exactly (\ref{z13}).

Symmetrizing (\ref{z13}) in lower
indices $\mu, \nu, \alpha, \sigma$ and $ \tau$ and
expressing $\Pi$ in terms of completely symmetric tensor $S$ \p{z12}
we obtain the identity
\be
S_{\rho(\mu\alpha}^{\ \ \ \ \gamma}S_{\sigma\tau\nu)}^{\ \ \ \ \rho}
-\delta^\gamma_{(\mu}S_{\alpha\sigma\tau\nu)}=0 \ .\label{z14}
\ee
It holds universally for all symmetric spaces
(\ref{one}). This identity plays fundamental role in the
construction of the quaternionic potentials for these spaces in
the next Section.

\section{Killing potentials and the Hamiltonian}

Now we use the results
of the previous section to find the Hamiltonians and
Killing potentials for the quaternionic symmetric spaces.
First of all we remark that the $Sp(1)$ invariance requires the
indices $i$ of $u^-_i$ or $q^{+i}$ in $\lpr(Q^+, q^+, u^-)$
to be contracted invariantly. The
only nonvanishing combination satisfying this condition is
$(q^+u^-)=q^{+i}u^-_i$. Therefore $\lpr$ depends only on $Q^+$ and
$(q^+u^-)$,
\be \lpr =\lpr (Q^+,(q^+u^-)) .\label{qu}
\ee
Since $\lpr $ has a $U(1)$-charge 4, the form \p{qu} implies
that $\lpr$ is homogeneous of degree 4 in $Q^+$. Moreover, since $Q^+$
is a remnant of coset coordinates corresponding to coset generators,
the point $Q^+=0$ must be nonsingular. Therefore $\lpr$ must be
proportional to a fourth order polynomial in $Q^+$.
Finally, since $\lpr$ is homogeneous of degree 2 in $Q^+,q^+$, the above
arguments restrict $\lpr$ to be
\be \lpr ={P^{+4}(Q^+)\over (q^+u^-)^2} \label{formofl}\ee
with a fourth order polynomial $P^{+4}$.
We are going to prove that
\be P^{+4}(Q^+)=\xi \; S_{\alpha\beta\gamma\delta}\;
Q^{+\alpha}Q^{+\beta}Q^{+\gamma}Q^{+\delta}\label{P^4}
\ee
with a constant $\xi$ to be defined later.

As explained in the Introduction, our proof consists of giving
explicit expressions for
the Killing potentials. We take
\be
Sp(1):\;\;\;\; K^{++}_{ij}=2(q^+_iq^+_j-u^-_iu^-_j\lpr), \label{su2curr}
\ee\be
H:\;\;\;\; K^{++}_a=t_{a\alpha\beta}Q^{+\alpha}Q^{+\beta},\label{Hcurr}
\ee\be
G/H\times Sp(1):\;\;\;\;K^{++}_{i\alpha}=2q^+_iQ^+_\alpha -
u^-_i(q^+u^-) \pa^-_\alpha\lpr\ ,\label{cosetcurr}
\ee
where $\pa^-_\alpha$ is the partial derivative with respect to
$Q^{+\alpha}$, and $t_{a\a\b}=\O_{\b\g}t_{a\a}^{\ \ \ \g}$.
It is straightforward to show that the $Sp(1)$ potential $K^{++}_{ij}$ is
conserved for an arbitrary polynomial $P^{+4}(Q^+)$.
The conservation law for
$K^{++}_a$ is satisfied for $P^{+4}$ given by (\ref{P^4}) but
still does not fix $\xi$. (With
the choice \p{Hcurr} the fourth order polynomial  $P^{+4}$ \p{P^4} is
proportional to $g^{ab}K_a^{++}K_b^{++}$.
Therefore it is similar to the second order Casimir
operator for the group
$H$ and this makes the conservation of $K^{++}_a$ obvious.)

A Killing potential $K^{++}$ induces the transformation
\be
\delta Q^{+\alpha} \sim \{Q^{+\alpha}, K^{++}\}
\ee
on the coordinates $Q^+$.
The origin $Q^{+\a}=0$ is stable under the transformations induced by
the Killing potentials $K^{++}_{ij}$ and $K^{++}_a$ but not
$K^{++}_{i\a}$ which shows that the stability subgroup is precisely
$H\times Sp(1)$.

Finally, a direct calculation shows that the
conservation law for the coset potential is reduced to the
equation
\be
Q^+_\alpha P^{+4} -{1\over 2}\{\pa_\alpha^-P^{+4} , P^{+4}\}^{--}=0\ .
\label{16}\ee
It imposes a nontrivial restriction on the structure constants
$t_{a\alpha}^{\ \ \ \beta}$ and $t_{\alpha\beta}^{\ \ \ a}$.
Substituting \p{P^4} into \p{16} we obtain an algebraic
equation for $S$,
\be
\left[ \Omega_{\alpha\mu}\; S_{\nu\rho\sigma\gamma} -
12\; \xi \; S_{\alpha\mu\nu\tau}
S_{\rho\sigma\gamma}^{\ \ \ \ \ \tau}\right]
Q^{+\mu} Q^{+\nu} Q^{+\rho} Q^{+\sigma} Q^{+\gamma} =0.
\label{sought}\ee
Comparing this with eq. \p{z14}  we find $\xi=1/12$.  Then
eq-ns \p{formofl}, \p{P^4} together with \p{anti} fix the Hamiltonian to
the form \p{l}.
One can similarly express $P^{+4}$ in terms of the coset
Killing potentials, or the $Sp(1)$ Killing potentials:
\be P^{+4}=-\frac{1}{16}\e^{ij}\O^{\a\b}K^{++}_{i\a}
K^{++}_{j\b}=-\frac{1}{8}K^{++ij}K^{++}_{ij}\ .\ee

 It is straightforward to check
that the Killing potentials given in \p{su2curr}, \p{Hcurr} and
\p{cosetcurr} generate precisely the  algebra \p{2.5}--\p{2.9}.

In summary,  we have found the harmonic potentials for the quaternionic
symmetric spaces. They turned out to be simple fourth order polynomials in
$Q^+$ similar to the quadratic Casimir operator of the subgroup $H$.
It seems plausible that a
more general class of homogeneous nonsymmetric
quaternionic spaces of Alekseevskii \cite{a16}
can be described in the same way.

\vskip 1.5cm
{\bf Acknowledgements}. We are grateful to E. Ivanov and V. Ogievetsky
for their interest and valuable discussions.  A.G. acknowledges
support by the U.S. National Science Foundation, grant PHY-90096198.
O.O. would like to thank the Particle Physics Group of the Johns Hopkins
University for hospitality during the course of the work.


\begin{thebibliography}{99}
\bibitem{egh} T. Eguchi, P.B. Gilkey and A.J. Hanson, {\sl Phys. Reports}
66 (1980) 213.
\bibitem{baggwitt} J. Bagger and E. Witten, {\sl Nucl. Phys. B}
222 (1983) 1.
\bibitem{bagger} J.A. Bagger, A.S. Galperin, E.A. Ivanov and V.I. Ogievetsky,
{\sl Nucl. Phys. B} 303 (1988) 522.
\bibitem{gikos} A. Galperin,
E. Ivanov, S. Kalitzin, V. Ogievetsky and E. Sokatchev,
{\sl Class. Quantum Grav.} 1 (1984) 469.
\bibitem{gio} A. Galperin, E. Ivanov and O. Ogievetsky, {\it Harmonic Space
Description of Quaternionic Manifolds}, preprint JHU-TIPAC-920023,
MPI-Ph/92-85, ENSLAPP-L-405/992.
\bibitem{zu} B. Zumino, {\sl Phys. Lett. B} 87 (1979) 203.
\bibitem{go} A. Galperin and V. Ogievetsky, {\sl Class. Quantum Grav.}
8 (1991), 1757.
\bibitem{wolf} J. Wolf, {\sl J. Math. Mech.} 14 (1965) 1033.
\bibitem{a16} D.A. Alekseevskii, {\it Math. USSR Izvestija} 9 (1975) 297.
\end{thebibliography}
\end{document}